\documentclass[aps,pra,groupedaddress,twocolumn,showpacs]{revtex4}

\usepackage{epsfig}
\usepackage{color}
\begin{document}

\title{Quantum coherence and population trapping in three-photon
processes}

\author{Caroline Champenois}
\email{caroline.champenois@up.univ-mrs.fr} \affiliation{Physique
des Interactions Ioniques et Mol\'eculaires (CNRS UMR 6633),
Universit\'e de Provence, Centre de Saint J\'er\^ome, Case C21,
13397 Marseille Cedex 20, France} \author{Giovanna Morigi}
\affiliation{Grup d'Optica, Departament de Fisica, Universitat
Autonoma de Barcelona, 08193 Bellaterra, Spain} \author{J\"urgen
Eschner} \affiliation{ICFO - Institut de Ci{\`e}ncies
Fot{\`o}niques, Mediterranean Technology Park, 08860 Castelldefels
(Barcelona), Spain}

\date{\today}

\begin{abstract} The spectroscopic properties of a single, tightly
trapped atom are studied, when the electronic levels are coupled
by three laser fields in an $N$-shaped configuration of levels,
whereby a $\Lambda$-type level system is weakly coupled to a
metastable state. We show that depending on the laser frequencies
the response can be tuned from coherent population trapping at
two-photon resonance to novel behaviour at three photon resonance,
where the metastable state can get almost unit occupation in a
wide range of parameters. For certain parameter regimes the system
switches spontaneously between dissipative and coherent dynamics
over long time scales. \end{abstract}


\maketitle

\section{introduction} Atomic coherence has been demonstrated to
be an efficient tool for achieving control of the interaction
between electromagnetic fields and an atomic sample. It is at the
basis, for instance, of the realization of quantum non-linear
optical devices~\cite{braje04, balic05, eisaman05}, of quantum
phase gates~\cite{lukin00, ottaviani03, friedler05}, atomic
transistors~\cite{micheli04} and of high-precision measurement
techniques~\cite{santra05}.

A paradigmatic system exhibiting atomic coherence effects is the
so-called $\Lambda$-configuration of atomic levels, where two
(meta)stable states are coupled by two light fields to the same
excited state. When the coupling happens at two-photon resonance,
the system exhibits Coherent Population Trapping, by forming an
atomic coherence between the two stable states which decouples
from the radiation \cite{arimondo96}. More complex configurations
of levels offer richer dynamics, whose understanding is relevant
for applications of coherent control of complex
systems~\cite{agarwal96, phaseonium, morigi02}. The interpretation
of their dynamics is often non-trivial, yet in some parameter
regimes analogies may be found with simpler level systems which
are better understood. This usually helps developing tools for
controlling and manipulating the quantum dynamics of the more
complex system through external parameters~\cite{agarwal96,
hemmer00, lukin03}.

\begin{figure}[htb]
\begin{center}
\epsfig{file=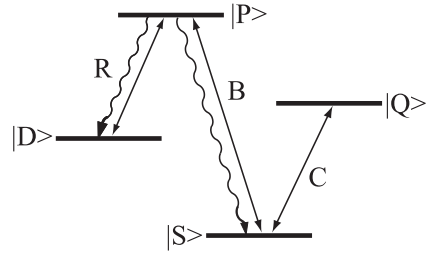, height=0.4\hsize} \caption{$N$ Level
scheme: The states $|D\rangle$, $|P\rangle$, $|S\rangle$ form a
$\Lambda$- configuration; additionally, state $|S\rangle$ couples
weakly to the metastable state $|Q\rangle$. The solid lines and
letters $R,B,C$ indicate the laser couplings, the wavy lines the
radiative decay channels. Parameters and possible atomic species
are discussed in the text.} \label{fig_N}
\end{center}
\end{figure}

In this work we study how the dynamics of a $\Lambda$-system is
modified by an additional coupling of one of the stable states to
a fourth, metastable state, as depicted in Fig.~\ref{fig_N}. Due
to its shape, we denote this configuration as $N$-level scheme. We
show that even weak coupling to the fourth level gives rise to
strong modification of the dynamics whenever the three-photon
resonance condition between the outer levels is fulfilled. In
particular, in certain parameter regimes the fourth state exhibits
quasi-unit occupation probability; in other situations the
behaviour switches from dissipative transient to coherent
asymptotic dynamics.

Our analysis applies to alkali-like atomic species, as well as to
some alkaline-earth elements. We consider in particular the case
of a single $^{40}$Ca$^{+}$ ion in a radiofrequency trap. For this
system we also study the effect of the oscillatory motion on the
dynamical behaviour.

$N$-level schemes have been extensively studied
in~\cite{taichenachev99, goren03, goren04} in the framework of
Electromagnetically Induced Absorption~\cite{akulshin98}. A
peculiar difference of those systems with the scheme we
investigate here is the stability of the fourth level, which
critically affects the response of the system. As a result, the
narrow lines at three-photon resonance, which we report in this
manuscript, cannot be explained in terms of transfer of
coherence~\cite{taichenachev99, goren03}, but are instead
intimately related to coherent population trapping, as we will
argue. Coherent population trapping and dark resonances have been
discussed in~\cite{lukin99, yelin03} for a configuration similar
to the one we discuss here, with the important difference that
in~\cite{lukin99, yelin03} the unstable state can decay in all
three stable states. In this latter system, Doppler-free-like
absorption resonances \cite{ye02} and three-photon
electromagnetically induced transparency (EIT) \cite{zibrov02}
have been observed. We will comment on how our model system
reproduces and differs from these studies. Finally, three-photon
resonances have been studied in connection with
metrology~\cite{appasamy95, hong05}. Indeed, the type of system we
consider is encountered in atomic clocks, where transition
$|S\rangle\to |Q\rangle$ is the clock transition of, say, a
$^{40}$Ca$^+$ ion or a $^{87}$Sr atom, and the dynamics we predict
may have applications for high-precision optical clocks. In this
context, we also study how the oscillatory motion inside an ion
trap modifies the spectroscopic signals.

This article is organized as follows. In Sec.~\ref{model} the
theoretical model is introduced, in Sec.~\ref{immobile} the
theoretical analysis and predictions are reported, and in
Sec.~\ref{motion} we extend it when the oscillatory motion of a
trapped atom is considered. In Sec.~\ref{Sec:Conclusions}
discussions and conclusions are reported, and in the appendix the
model at the basis of calculations in Sec.~\ref{motion} is
described.

\section{Theoretical model} \label{model}

The atomic system we consider is composed by 4 electronic levels
which are coupled by laser fields, according to the $N$-shaped
scheme depicted in Fig.~\ref{fig_N}. Here, states $|S\rangle$,
$|D\rangle$ and $|Q\rangle$ are (meta)stable, state $|P\rangle$ is
unstable and decays radiatively into $|S\rangle$ and $|D\rangle$.
This configuration of levels is realized, for instance, in
alkaline-earth atoms with hyperfine structure and in alkali-like
ions with a metastable $d$-orbital, such as Hg$^+$, Ba$^+$,
Sr$^+$, or Ca$^+$. In this manuscript we make reference to the
typical parameters of a $^{40}$Ca$^+$ ion. In this case, the
relevant levels can be identified with the states
$|S\rangle=|S_{1/2}\rangle$, $|P\rangle=|P_{1/2}\rangle$,
$|D\rangle=|D_{3/2}\rangle$, and $|Q\rangle=|D_{5/2}\rangle$.
Here, the transition $|S\rangle\to|Q\rangle$, marked by C in
Fig.~\ref{fig_N}, is an electric quadrupole transition with a
linewidth of the order of 0.1~Hz, while $|S\rangle$ and
$|D\rangle$ couple to the excited state $|P\rangle$ with electric
dipole transitions (marked by B and R, respectively). We denote by
$\omega_{ij}$ ($i,j=S,P,D,Q$) the resonance frequencies of the
transitions. Transitions $|S\rangle\to |P\rangle$ and
$|D\rangle\to |P\rangle$ are coupled by lasers at frequency
$\omega_B$ and $\omega_R$ and Rabi frequencies $\Omega_B$ and
$\Omega_R$, respectively. These three states form a
$\Lambda$-configuration of levels. Transition
$|S\rangle\to|Q\rangle$ is driven by a laser at frequency
$\omega_C$ and Rabi frequency $\Omega_C$.

We denote by $\rho$ the density matrix for the ion's internal
degrees of freedom, while we treat the center of mass variables
classically. We denote by $x(t)$ the time-dependent position of
the atom. The master equation for the density matrix $\rho$ is
\begin{equation} \label{Master:Eq} \frac{\partial}{\partial
t}\rho=-\frac{\rm i}{\hbar}[H,\rho]+{\cal L}\rho \end{equation}
where Hamiltonian $H$ gives the coherent dynamics and is
decomposed into the terms $$H=H_0+H_I$$ where \begin{eqnarray} H_0
&=&-\hbar\Delta_B|P\rangle\langle P|-\hbar\Delta_C|Q\rangle\langle
Q| +\hbar(\Delta_R-\Delta_B)|D\rangle\langle D|
 \end{eqnarray}
gives the internal energies in the reference frames of the lasers,
with detunings defined as $\Delta_B=\omega_B-\omega_{PS}$,
$\Delta_R=\omega_R-\omega_{PD}$, and
$\Delta_C=\omega_C-\omega_{QS}$, while
 \begin{eqnarray} \label{Ham}H_I
&=&\frac{\hbar\Omega_B}{2}{\rm e}^{{\rm i}k_Bx(t)}|P\rangle\langle
S|+\frac{\hbar\Omega_R}{2}{\rm e}^{{\rm i}k_Rx(t)}|P\rangle\langle D|\\
& & +\frac{\hbar\Omega_C}{2}{\rm e}^{{\rm
i}k_Cx(t)}|Q\rangle\langle S|+{\rm H.c.} \nonumber \end{eqnarray}
gives the laser coupling, where $\Omega_j$ denote the Rabi
frequencies and $k_j$ the corresponding laser wave vectors
($j=B,R,C$).

The relaxation operator \begin{eqnarray} {\cal L}\rho
&=&-\frac{1}{2}\gamma_P \left(\rho|P\rangle\langle
P|+|P\rangle\langle P|\rho\right)\\
& &+\beta_{PS}\gamma_P|S\rangle\langle P|\rho|P\rangle\langle S|
+\beta_{PD}\gamma_P|D\rangle\langle P|\rho|P\rangle\langle D|
\nonumber \end{eqnarray} describes the radiative processes,
coupling $|P\rangle$ to states $|S\rangle$ and $|D\rangle$, with
branching ratio $\beta_{PS}/\beta_{PD}\simeq 15$ for Ca$^{+}$ and
$\beta_{PS}+\beta_{PD}=1$.

The radiative decay of state $|Q\rangle$, whose lifetime is about
1~s for Ca$^{+}$, will be neglected in the analytical model we
present below, but it is taken into account in the numerical
calculations. It should also be noted that there is a large
difference, by some orders of magnitude, between the dipole and
the quadrupole couplings. In this manuscript we will focus on
situations in which state $|Q\rangle$ is weakly coupled to the
$\Lambda$-scheme, and which are thus experimentally feasible with
standard laser sources. Finally, we will make reference to the
dynamics of a {\it single} ion, like it can be realized in
radio-frequency traps, and we will characterize its response by
means of the occupation probabilities of the atomic levels, which
can be monitored by resonance fluorescence or electron shelving
techniques \cite{dehmelt75}. We will also take into account the
ion's oscillatory motion in the trap.

\section{Internal dynamics of a localized
particle\label{immobile}}

In this section we focus on the solutions of Eq.~(\ref{Master:Eq})
when the motion of the particle can be neglected, i.e. for steep
traps and efficient cooling, such that the amplitude of its
residual oscillations is much smaller than the laser wavelength.
We study the dynamics with simple analytical models and compare
their predictions with the results obtained from numerically
solving the optical Bloch equations, derived from
Eq.~(\ref{Master:Eq}). We analyze the steady state and the time
evolution under two particular conditions: {\it (i)} the
three-photon resonance case,
\begin{equation}
\label{3-photon}\Delta_B-\Delta_R-\Delta_C=0 \ \ {\rm and }\ \
\Delta_C \ne 0
\end{equation}
in which states $|D\rangle$ and
$|Q\rangle$ are resonantly coupled by three-photon processes, and
{\it (ii)} the two+one-photon resonance case,
\begin{equation}
\label{2-photon}\Delta_R-\Delta_B=0 \ \ {\rm and }\ \ \Delta_C=0
\end{equation}
in which states $|S\rangle$ and $|D\rangle$ are coupled resonantly by a
two-photon transition while $|S\rangle$ and $|Q\rangle$ are coupled resonantly
by a one-photon process.

The major difference between the two cases is that when the
two+one-photon resonance condition {\it (ii)} is fulfilled, the
steady state response of the $\Lambda$-system alone would be
characterized by a dark resonance, or coherent population
trapping, resulting from the destructive interference between the
two excitation paths $|S\rangle \to |P\rangle$ and $|D\rangle \to
|P\rangle$ \cite{arimondo96}. This has profound consequences also
for the four-level dynamics, as will be shown below.

\subsection{Dressed states analysis\label{dressing}}

\begin{figure}[htb]
\begin{center}
\epsfig{file=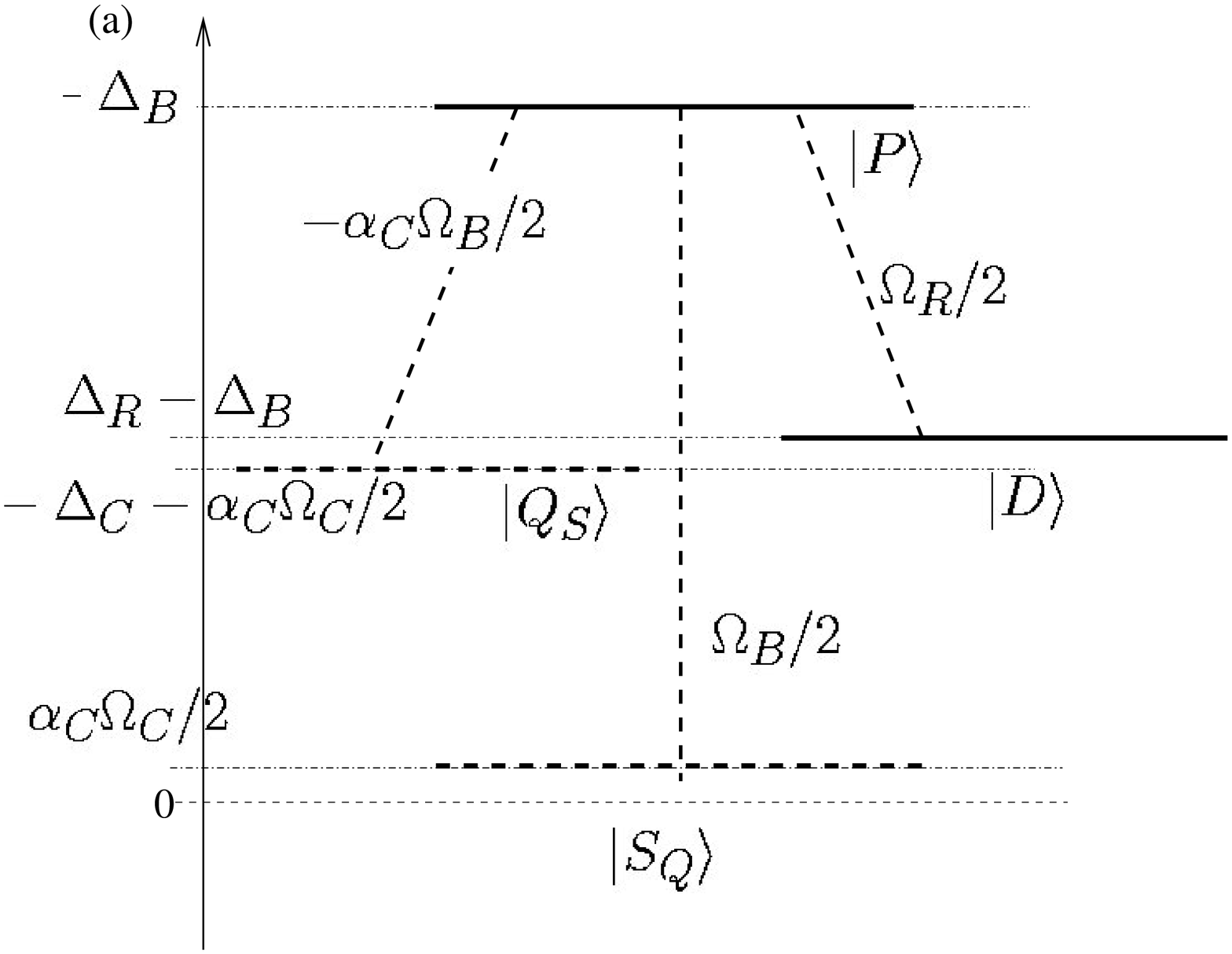, width=0.9\hsize}\\
\epsfig{file=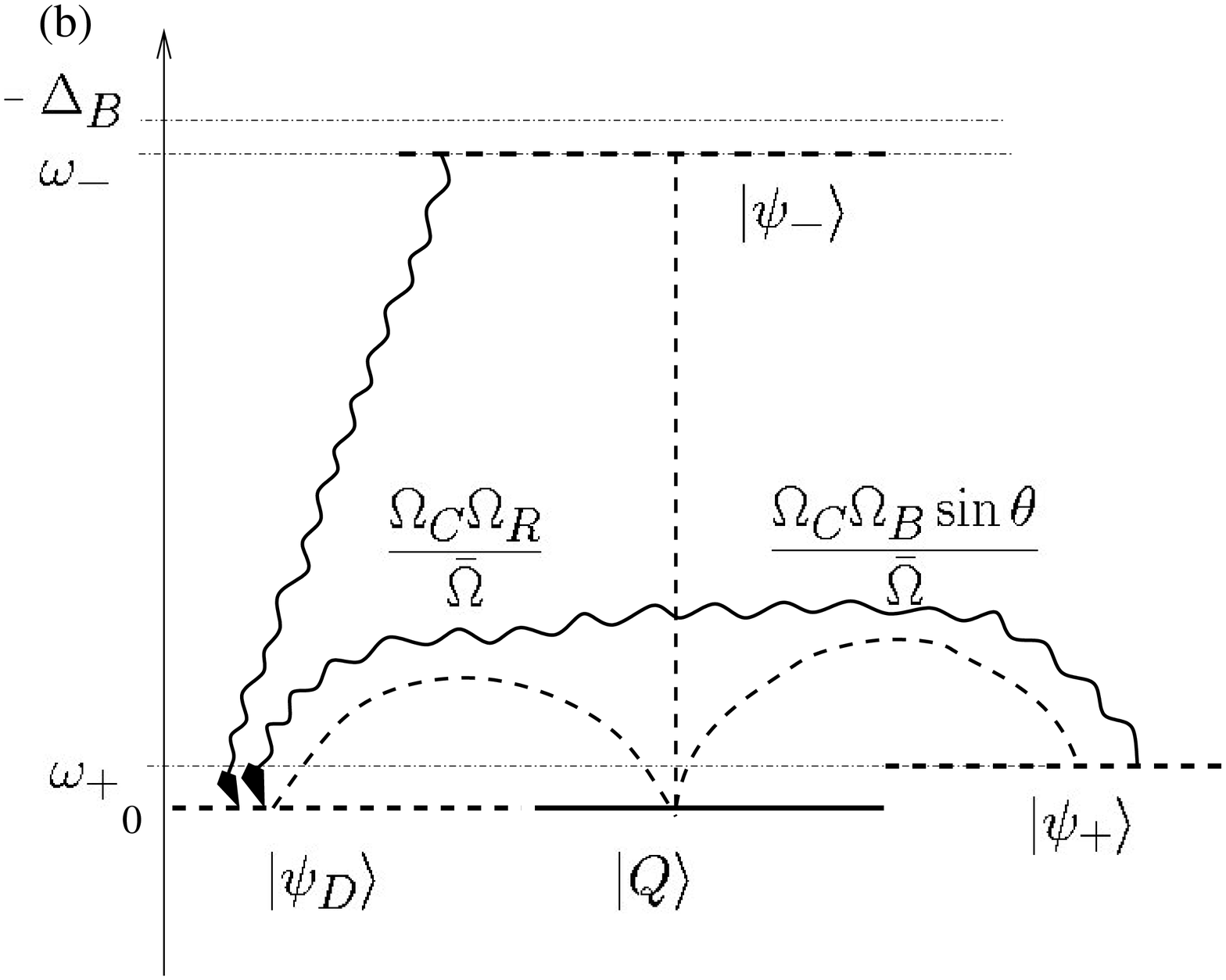, width=0.9\hsize}
 \caption{Dressed states picture for (a) the three-photon resonance case and (b) the two+one-photon
 resonance case. See text for parameters and definitions.
  \label{fig_dressed}}
\end{center}
 \end{figure}

In order to get some insight, we evaluate the dressed states of
the system in the two limiting cases. We first focus on the
three-photon resonance, condition~(\ref{3-photon}), in the
situation when the coupling between states $|Q\rangle$ and
$|S\rangle$ can be treated perturbatively. We hence assume
$\Delta_C\neq 0$ and $\Omega_C \ll |\Delta_C|$. The coupling between
$|Q\rangle$ and $|S\rangle$ is in first order in the perturbation
parameter $\alpha_C=\Omega_C/2\Delta_C$, and corrections to the
states $|Q\rangle$ and $|S\rangle$ are at second order in
$\alpha_C$, according to
\begin{eqnarray}
\left|S_Q\right> & =&{\mathcal N} \left(|S\rangle+\alpha_C|Q\rangle\right)\\
\left|Q_S\right> & =&{\mathcal N}
\left(|Q\rangle-\alpha_C|S\rangle\right) \end{eqnarray} where
${\mathcal N}$ gives the correct normalization. The
eigenfrequencies for these two states are $\alpha_C\Omega_C/2$ and
$-\Delta_C-\alpha_C\Omega_C/2$, as displayed in Figure
\ref{fig_dressed}(a). Here, $|D\rangle$ and $|Q_S\rangle$ are
resonantly coupled by an effective two-photon process, and the
system can be pumped into the eigenstate \begin{equation}
|\Psi_{NC}\rangle ={\cal N}'\left({\cal E}
|D\rangle+|Q_S\rangle\right) \label{psiNC} \end{equation} with
$${\cal E}=\frac{\Omega_B}{\Omega_R}\alpha_C$$ and normalization
factor ${\cal N}'$. This state is stable  at second order in
$\alpha_C$, it has the property of a dark state which is occupied
asymptotically, thus signalling coherent population
trapping~\cite{arimondo96}. According to this description, the
corresponding electronic occupations at steady state, ${\cal
P}_j^{(NC)}=|\langle j|\Psi_{NC}\rangle|^2$ ($j=Q,D,S,P$), are
\begin{eqnarray}
{\cal P}_Q^{(NC)} & = & \frac{1}{1+\alpha_C^2+{\cal E}^2}+{\rm O}(\alpha_C^4,{\cal E}^4 )\label{popQ}\\
{\cal P}_D^{(NC)} & =& \frac{{\cal E}^2}{1+{\cal E}^2}+{\rm O}(\alpha_C^4,{\cal E}^4)\\
{\cal P}_S^{(NC)}  & = & \frac{\alpha_C^2}{1+\alpha_C^2+{\cal E}^2} +{\rm
O}(\alpha_C^4,{\cal E}^4)
 \end{eqnarray}
while ${\cal P}_P^{(NC)}={\rm O}(\alpha_C^4)$. Hence, the
parameter ${\cal E}$, or more precisely the ratio
$\Omega_B/\Omega_R$ compared to $1/\alpha_C$, determines the
distribution of population between states $|Q\rangle$ and
$|D\rangle$. A typical experimental situation is that $\Omega_B$
and $\Omega_R$ are similar, such that we concentrate on the case
$\Omega_B/\Omega_R \ll 1/\alpha_C$, i.e. ${\cal E} \ll 1$, and we
see that the atom will occupy $|Q\rangle$ with almost unit
probability. In this regime, the linewidth of state $|Q\rangle$ is
due to higher-order coupling in $\alpha_C$ to state $|P\rangle$,
and scales with $(\alpha_C^2\Omega_B/\Omega_R)^2$.

We now consider the case in which the two+one-photon resonance condition
(\ref{2-photon}) is fulfilled.  We still restrict the discussion to the regime
in which $\Omega_C$ is weak compared to all other coupling terms. In this case
it is convenient to consider the eigenstates of the $\Lambda$ system formed by
$\{ |S\rangle, |P\rangle, |D\rangle \}$,
\begin{eqnarray*}
&&|\psi_D\rangle=(\Omega_R|S\rangle-\Omega_B|D\rangle)/\bar{\Omega} \label{darkstate}\\
&&|\psi_{+}\rangle=\cos\theta|P\rangle+\sin\theta(\Omega_B|S\rangle+\Omega_R|D\rangle)/\bar{\Omega}\\
&&|\psi_{-}\rangle=-\sin\theta|P\rangle+\cos\theta(\Omega_B|S\rangle+\Omega_R|D\rangle)/\bar{\Omega}
\end{eqnarray*} with $\bar{\Omega}=\sqrt{\Omega_R^2+\Omega_B^2}$
and $\tan
\theta=(\Delta_B+\sqrt{\Delta_B^2+\bar{\Omega}^2})/\bar{\Omega}$
with $0 \leq \theta < \pi/2$. The dressed states of the
diagonalised $\Lambda$-system are at frequencies $$\omega_D=0$$
$${\rm and}~~~\omega_{\pm} = -\frac{1}{2} \left(
\Delta_B\mp\sqrt{\Delta_B^2+\bar{\Omega}^2} \right)~.$$ The level
scheme in this new basis is sketched in Figure
\ref{fig_dressed}(b). Here, $|\psi_D\rangle$ is the dark state of
the $\Lambda$-system~\cite{arimondo96}. At two+one-photon
resonance, state $|Q\rangle$ is resonantly coupled to the dark
state $|\psi_D\rangle$ at rate
$\tilde{\Omega}_C=\Omega_C\Omega_R/\bar{\Omega}$, and for
sufficiently weak coupling, $\Omega_C\ll|\omega_{\pm}|$, there
will be a time scale on which the dynamics of the system can be
reduced to resonant two-level dynamics between these two states.
For longer times, off-resonant coupling between $|Q\rangle$ and
$|\psi_{\pm}\rangle$ gives rise to damping, and the system
approaches the steady state at a rate which scales with the ratio
$\Omega_B^2/\bar{\Omega}^2$. This damping gives rise also to the
small residual linewidth of the two+one-photon resonance in the
spectra shown in the next section.

\subsection{Steady state\label{steady_state}}

In order to illustrate the spectroscopic significance of the
dressed states, we now discuss the steady-state populations ${\cal
P}_j$ of the electronic levels ($j=S,P,D,Q$) as a function of
$\Delta_R$, for the two cases $\Delta_C\neq 0$ and $\Delta_C=0$.
For highlighting the peculiarities of the 4-level dynamics, we
compare them to the behaviour of the unperturbed $\Lambda$-system
($\Omega_C=0$), whose stationary level occupations are indicated
by the dashed curves; they exhibit the dark resonance at
$\Delta_R=\Delta_B$, corresponding to suppressed population of
$|P\rangle$. The effect of the weak coupling to state $|Q\rangle$
is represented by the solid curves.

\begin{figure}[htb]
\begin{center}
\epsfig{file=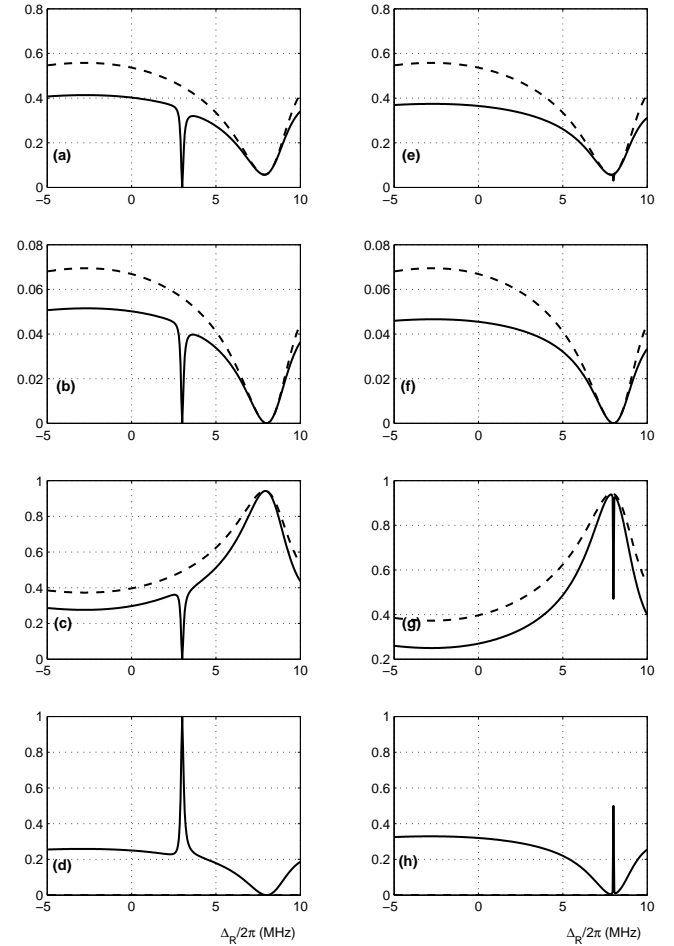, width=0.99\hsize}
 \caption{Steady state populations ${\cal P}_j$ as a function of
the detuning  $\Delta_R/2\pi$. Displayed are ${\cal P}_S$, ${\cal
P}_P$, ${\cal P}_D$ and ${\cal P}_Q$, from top to bottom. The
parameters are $\Delta_B=2\pi \times 8$ MHz, $\Omega_C=2\pi \times
0.05$ MHz, $\Omega_B=2\pi \times 10$ MHz, and $\Omega_R=2\pi
\times 2.5$ MHz. In (a)-(d) $\Delta_C=2\pi \times 5$ MHz, such
that the three-photon resonance is fulfilled at
$\Delta_R=2\pi\times 3$~MHz; in (e)-(g) $\Delta_C=0$, such that
the two+one-photon resonance is found at $\Delta_R=2\pi \times 8$
MHz. The dashed lines give the steady state populations for the
same parameters but no coupling to $|Q\rangle$ ($\Omega_C=0$).
\label{fig_Sstate}} \end{center} \end{figure}

Figure~\ref{fig_Sstate}(a)-(d) displays the case $\Delta_C\neq 0$.
One observes that coupling to $|Q\rangle$ does not change the
behaviour around the dark resonance, but it induces a critical
change when $\Delta_R$ is at three-photon resonance: at this value
all population is transferred to state $|Q\rangle$, Fig.
\ref{fig_Sstate}(d), while all other states are correspondingly
emptied. The width of this resonance is controlled by the ratio  $\Omega_R/\Omega_B$, as discussed in the previous section. With the parameters chosen in Fig.~\ref{fig_Sstate} ($\Omega_R/\Omega_B=0.25$), the resonance is narrow but it can be made broader by increasing this ratio.
In general, occupation of state $|Q\rangle$ at
three-photon resonance is controlled by the parameter ${\cal E}$,
as has been pointed out in the dressed state picture in
Sec.~\ref{dressing}. The important finding in this context is that
in the regime ${\cal E} \ll 1$, ${\cal P}_Q$ is very close to
unity, independent of the value of $\Omega_C$. This is only
limited for very small values of the coupling by the decay of
level $|Q\rangle$. In practical terms, this allows for robust
preparation of the atom in $|Q\rangle$, by tuning the lasers to
the three-photon resonance.

Figure \ref{fig_Sstate}(e)-(h) displays the stationary populations as a
function of $\Delta_R$ when $\Delta_C=0$. At two+one-photon resonance, the
coupling to $|Q\rangle$ gives rise to a transfer of 50\% of the atomic
population from the dark state of the $\Lambda$-system to $|Q\rangle$. In
contrast to the case of $\Delta_C \ne 0$, this is indeed the maximum occupation
that $|Q\rangle$ can achieve for $\Delta_C=0$. This is understood considering
the dressed state picture in Sec.~\ref{dressing}: for $\Delta_C=0$ dark state
and $|Q\rangle$ form a resonantly coupled two-level system, where damping is
weak and arises only from off-resonant coupling. Hence, at steady state the
populations of the two states are the stationary populations of a saturated
dipole.

The two examples reported here show that a weak coupling to a
fourth state which is metastable can change dramatically the
response of a $\Lambda$-system when the detunings fulfill, or are
around, the three-photon resonance condition. Three-photon
processes have been previously studied in \cite{lukin99, yelin03}
in an atomic model system where decay of state $|P\rangle$ into
$|Q\rangle$ is allowed. This constitutes a major difference to the
dynamics discussed here: in the model of~\cite{lukin99, yelin03}
one does not observe the narrow three-photon resonance for weak
coupling $\Omega_C$, as population is optically pumped into
$|Q\rangle$ for a wide range of values of $\Delta_R$.

\subsection{Time evolution at three-photon resonance
\label{evolution}}

\begin{figure}[htb]
\begin{center}
\epsfig{file=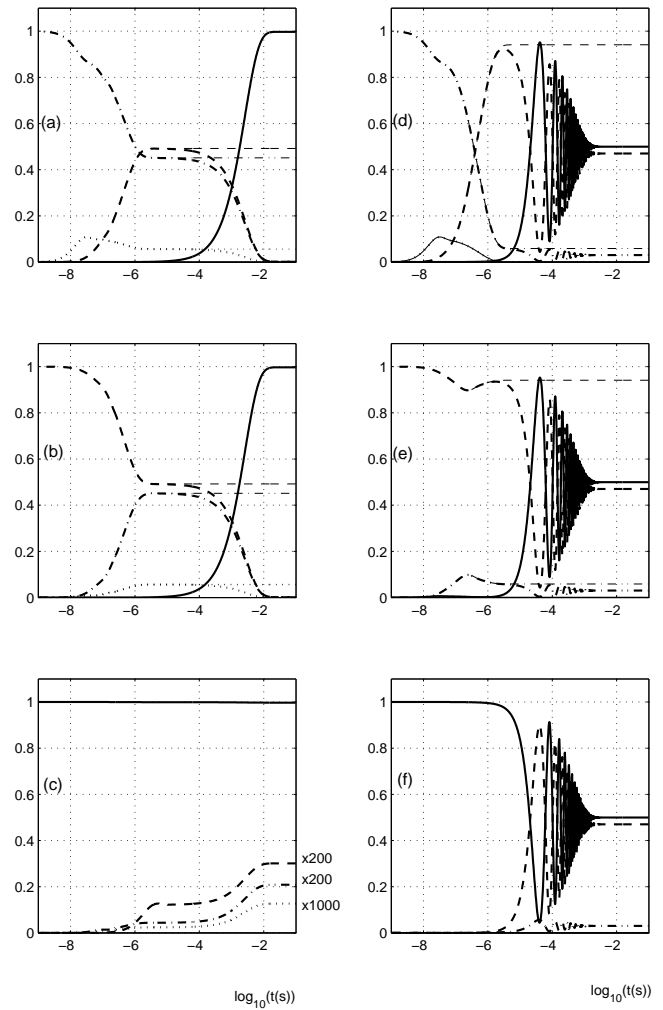, width=0.99\hsize} %
\caption{Time evolution of the atomic level populations at three-
(left) and two+one-photon resonance (right), for different initial
states. The lines correspond to the $|Q\rangle$ state (solid),
$|D\rangle$ state (dashed), $|S\rangle$ state (dash-dotted) and
$|P\rangle$ state (dotted). The initial state is $|S\rangle$ in
(a,d), $|D\rangle$ in (b,e) and $|Q\rangle$ in (c,f). Parameters
in (a-c) are the same as in Fig.~\ref{fig_Sstate}(a-d), and in
(d-f) the same as in Fig.~\ref{fig_Sstate}(e-h). The thinner
curves give the result in absence of coupling to state $|Q\rangle$
($\Omega_C=0$). \label{fig_evolution}} \end{center} \end{figure}

Let us now consider how the atomic level occupation evolves as a function of
time. Figure~\ref{fig_evolution}(a)-(c) displays the time evolution for
different initial conditions when the atom is driven at three-photon resonance
and when $|\Delta_C|\gg\Omega_C$, i.e. under conditions for which the atom is
found in $|Q\rangle$ at steady state. The thinner curves are plotted for
comparison and indicate the corresponding three-level dynamics, evaluated by
setting $\Omega_C=0$. When the initial state is $|S\rangle$ or $|D\rangle$
(Fig.~\ref{fig_evolution}(a-b)), one can identify a clear hierarchy of
couplings: on a short time scale, within about 1~$\mu$s for the chosen
parameters, the system evolves to the steady state of the
$\Lambda$-configuration. On a longer time scale ($\sim 1$~ms), population is
transferred to state $|Q\rangle$ through its coupling to the $\Lambda$-scheme.
When the system has been prepared in state $|Q\rangle$
(Fig.~\ref{fig_evolution}(c)), it essentially remains in that state during all
time, apart from a small redistribution of population from $|Q\rangle$ to
$|D\rangle$ on the slow time scale.

Such a dynamical behaviour indicates the appearance of quantum
jumps \cite{qjumps1,qjumps2,qjumps3}, i.e., of randomly
alternating phases of full and no fluorescence: after the emission
of a photon, which projects the atom into either $|S\rangle$ or
$|D\rangle$, the atom quickly assumes a quasi-steady state (the
steady state of the $\Lambda$-system) which has significant
population in $|P\rangle$ and is therefore likely to scatter more
photons. The average duration of these bright periods is given by
the slow time scale on which the system evolves into $|Q\rangle$.
With the parameters of this example, an average of about $3 \times
10^{3}$ photons are scattered during a bright period. When the
system has made a transition to $|Q\rangle$, signalled by a dark
time much longer than the typical interval between two scattered
photons \cite{nienhuis87}, then it will remain there for the
average duration of the dark periods, i.e. the long time scale on
which $|Q\rangle$ couples to the remaining three states
(Fig.~\ref{fig_evolution}(c)).

\begin{figure}[htb] \begin{center}
\epsfig{file=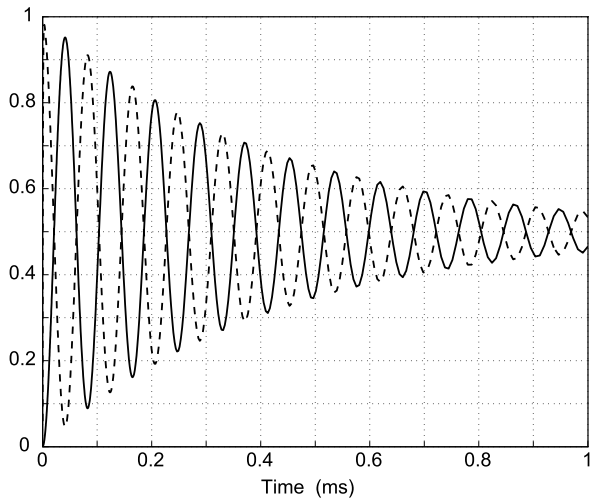, width=0.9\hsize} \caption{Rabi
oscillations between states $|\psi_D\rangle$ (dashed line) and
$|Q\rangle$ (solid line), corresponding to
Fig.~\ref{fig_evolution}(d) but plotted on a linear time scale.
\label{fig_Rabi}} \end{center} \end{figure}

Figure \ref{fig_evolution}(d)-(f) displays the time evolution out
of various initial states, when the two+one-photon resonance is
fulfilled. The dynamics are again well separated into different
time scales: the system accesses very quickly (within $\sim 1
\mu$s) the steady state of the $\Lambda$-system, which in this
case is the dark state of Eq.~(\ref{darkstate}) with no population
in $|P\rangle$. On the very long time scale ($>1$~ms), the global
steady state including $|Q\rangle$ is assumed. The important
observation, peculiar for this two+one-photon case, are
oscillations on the intermediate time scale, between the
populations of $|D\rangle$, $|S\rangle$ and $|Q\rangle$, whereby
$|D\rangle$ and $|S\rangle$ oscillate in phase with each other,
and in antiphase with $|Q\rangle$. This is consistent with the
dressed state analysis of Sec.~\ref{dressing} and corresponds to
Rabi oscillations between the dark state $|\psi_D\rangle$ and the
state $|Q\rangle$. Fig.~\ref{fig_Rabi} highlights these
oscillations, whose frequency is determined by the effective Rabi
coupling $\Omega_C\Omega_R/\bar{\Omega}$. Hence, the overall
dynamics of the two+one-photon resonance case are characterized by
an initial dissipative behaviour which evolves into a period of
coherent dynamics, i.e. Rabi oscillations between $|\psi_D\rangle$
and $|Q\rangle$; finally, these Rabi oscillations are also damped
out through off-resonant coupling to states $|\psi_{\pm}\rangle$,
which decay incoherently.

We remark that in this situation a single atom would be observed
to switch spontaneously between incoherent and coherent dynamics:
as the damping of the Rabi oscillations corresponds to the
emission of a photon, it will trigger the dynamics of
Fig.~\ref{fig_evolution}(d) to start again. This novel type of
quantum-jump behaviour should be revealed in an experiment through
long-time oscillations of the $g^{(2)}(\tau)$ photon-photon
correlation function.

Moreover, the coherent coupling between the dark and the
$|Q\rangle$ state can be brought to an extreme behaviour: in the
limit $\Omega_R\ll\Omega_B$ the dark state practically coincides
with state $|D\rangle$, and one observes Rabi oscillations between
states $|D\rangle$ and $|Q\rangle$, where population is directly
and coherently transferred between the two states by means of
three-photon processes.

\section{Effects of the center-of-mass motion\label{motion}}

So far we have neglected the effect of the center-of-mass motion on the atomic
dynamics. The motion may however critically affect the atomic response, and the
$N$-type level scheme we are considering has been subject of several studies of
how inhomogeneous broadening affects light transmission in atomic vapours
\cite{zibrov02, ye02, yelin03}. Let us start with some general considerations
for our particular system. In the dressed state picture at three-photon
resonance (condition~(\ref{3-photon})), motion of the atom at momentum
$\vec{p}$ gives rise to a Doppler effect which lifts the degeneracy between
states $|D\rangle$ and $|Q\rangle$. It thus gives rise to an instability of
state $|\Psi_{NC}\rangle$, Eq.~(\ref{psiNC}), which now couples to state
$|\Psi_C\rangle={\cal N}^{\prime}(|D\rangle-{\cal E}|Q_S\rangle)$ at rate
\begin{equation} {\cal R}=\langle\Psi_{NC}|H|\Psi_C\rangle\approx{\mathcal
E}\frac{\vec{p}}{m}\cdot(\vec{k}_R-\vec{k}_B+\vec{k}_C)
\end{equation}
This coupling may sensitively affect the dynamics of the system, due to the
narrow resonance condition. It vanishes, however, in the geometric Doppler-free
three-photon resonance condition (phase-matching condition)
$$\Delta\vec{k}=\vec{k}_R-\vec{k}_B+\vec{k}_C=\vec{0}.$$
This configuration has been studied in~\cite{yelin03,zibrov02}. In addition, in the
same configuration but for $\Delta\vec{k}\neq 0$, Doppler-insensitive
three-photon resonances have been observed when $|S\rangle\to|Q\rangle$ is in
the radio-frequency regime~\cite{zibrov02}. In this special case, the
radio-frequency coupling gives rise to sidebands and thus to a discrete
spectrum of excitations on the transition $|S\rangle\to|Q\rangle$ which couple
quasi-resonantly to $|D\rangle$ for different velocity classes~\cite{yelin03}.

In our model system, $^{40}$Ca$^{+}$, all transitions are in the
optical regime. We account for the oscillatory motion of the ion
inside the trapping potential by a time-dependent position
$$\vec{x}(t)=\vec{x}_0\cos\nu t,$$ where $\vec{x}_0$ is the
classical oscillation amplitude and $\nu$ the frequency of
oscillation \footnote{Both the driven (micro-) and the secular
(macro-) motion of a trapped ion may be described this way; unless
quantum effects in the macro-motion are relevant, the thermal
state of a laser cooled ion is modelled by additional integration
over a thermal distributions of oscillation amplitudes.}. In the
Hamiltonian~(\ref{Ham}), the effect is a modulation of the
radiative coupling. In this manuscript we will assume the
Lamb-Dicke regime, using the Lamb-Dicke parameters
$\eta_j=\vec{k}_j \cdot\vec{x}_0/2$ ($j=S,P,D,Q$) as small
perturbative parameters. We then use a Floquet ansatz for studying
the stationary response of the system. The basic equations are
reported in the appendix.

Figure \ref{fig_sideband} shows the steady state populations of
the four electronic levels around the three-photon resonance (left
column) and the two+one-photon resonance (right column) for
different laser beam geometries. The dashed curves correspond to
the case where the lasers are co-propagating, which for Ca$^{+}$
gives rise to a small three-photon Doppler effect:  the two metastable states have very close energy
levels, and in this configuration, the effective wavevector is
$\Delta\vec{k}=\vec{k}_B\times 0.003$. We see that, in this case,
the effect of the motion does not change appreciably the steady
state occupation.

\begin{figure}[htb] \begin{center}
\epsfig{file=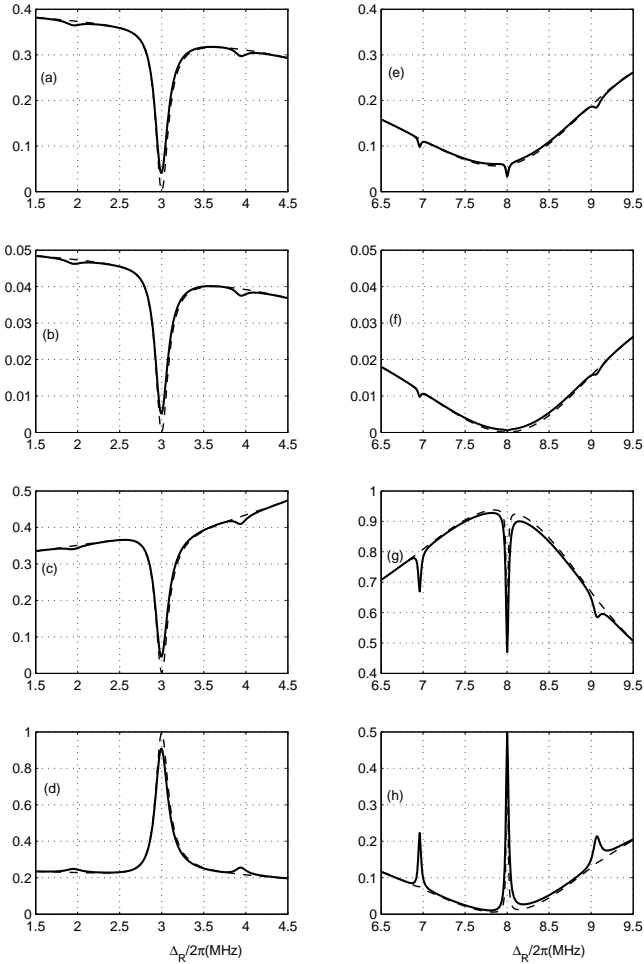, width=0.99\hsize}
 \caption{ Steady state populations ${\cal P}_S$, ${\cal P}_P$, ${\cal
P}_D$, and ${\cal P}_Q$ (from top to bottom) versus detuning
$\Delta_R/2\pi$ for the same parameters as in
Fig.~\ref{fig_Sstate} and taking into account the oscillation of
the Ca$^+$ ion in a trap at frequency $\nu=2\pi\times 1$~MHz and
Lamb-Dicke parameters $|\eta_{B,R,C}|=(0.1,0.046,0.054)$. The
dashed line corresponds to the case of co-propagating lasers
(minimal $\Delta \vec{k}=\vec{k}_B\times 0.003$), the solid line
to $\vec{k}_R$ and $ \vec{k}_C$ co-propagating against $
\vec{k}_B$ (maximal $\Delta \vec{k}\simeq \vec{k}_B\times 2 $).
The left column illustrates the three-photon resonance, the right
one the two+one-photon resonance.\label{fig_sideband}}
\end{center} \end{figure}

\begin{figure}[htb] \begin{center}
\epsfig{file=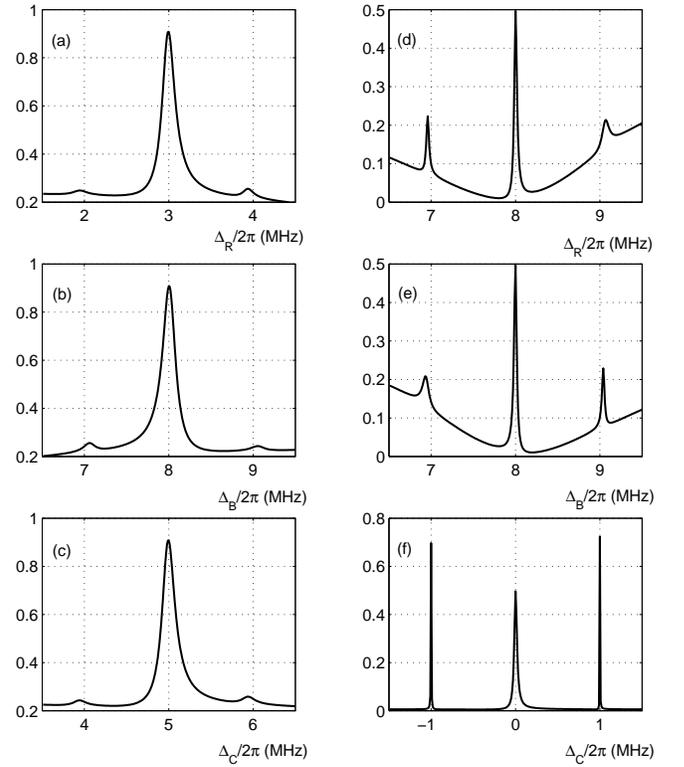, width=0.99\hsize}
\caption{Steady state population ${\cal P}_Q$ versus detunings
$\Delta_R/2\pi$, $\Delta_B/2\pi$, and $\Delta_C/2\pi$ (from top to
bottom), for the same parameters as in Fig.~\ref{fig_sideband},
and for $\vec{k}_R$ and $ \vec{k}_C$ copropagating against $
\vec{k}_B$ ($\Delta \vec{k}\simeq \vec{k}_B\times 2 $). The left
column illustrates the three-photon resonance, the right one the
two+one-photon resonance. \label{fig_sideband2}} \end{center}
\end{figure}

The solid lines in Figure~\ref{fig_sideband} correspond to a
Doppler sensitive geometry, where $\vec{k}_B$ counter-propagates
against $\vec{k}_R$ and $ \vec{k}_C$. For this geometry one can
observe sidebands in the resonance profile. Moreover, there are
major differences between the three-photon and the two+one-photon
resonance cases, in which the sideband signals are significantly
narrower and higher. One also observes that in the three-photon
resonance case the height of the central band of ${\cal P}_Q$ is
reduced by the effect of the motion to a value smaller than unity
(Fig.~\ref{fig_sideband}(d)) while for the two+one-photon
resonance the central band of ${\cal P}_Q$ still reaches the
maximum value 1/2. Further insight can be gained by comparing the
spectra obtained by scanning the three different lasers, as shown
in Fig.~\ref{fig_sideband2}. One clearly observes that the
sidebands are especially high when they fall in the dark-resonance
region, and that they may be larger than the central band when the
two+one-resonance condition is fulfilled between the dark state
and one of the sidebands (Fig.~\ref{fig_sideband2}(f)). In this
latter case, they achieve a value larger than 1/2 due to the
contribution of three-photon processes which include one sideband
transition.

\section{Discussion and conclusions}\label{Sec:Conclusions}

The weak coupling of a $\Lambda$-shaped 3-level system to a fourth
metastable state modifies critically the dynamics at three-photon
resonance. In a wide regime of parameters, the metastable level is
occupied with unit probability at steady state. The time evolution
shows that the dynamics of the system is characterized by two time
scales, a short one corresponding to the dynamics of the
$\Lambda$-system alone, and a longer one corresponding to its
coupling to the metastable state: on this time scale population is
transferred into it. When the $\Lambda$-system is driven at
two-photon resonance and the three-photon resonance condition is
fulfilled, the system first accesses the 3-level dark state
through dissipation and then switches to coherent dynamics,
characterized by Rabi oscillations between dark and metastable
state. By choosing the coupling properly, these dynamics can
reduce to direct coherent transfer (Rabi oscillations) between the
two extremal states coupled at three-photon resonance.

In general, the center-of-mass motion  modifies substantially the
response of the system, due to the sensitivity of the narrow
resonances to Doppler shifts. The narrow three-photon resonance is
recovered in configurations of the laser beams for which the
Doppler effect on the three-photon transition is suppressed. In
this manuscript we have considered the effect of the oscillatory
motion of a tightly trapped atom, and we have observed that the
motional sidebands which appear in the spectra can be
significantly enhanced due to interfering two- and three-photon
processes.

In an experimental realization, finite laser bandwidths will be
detrimental for the creation of quantum coherences. The
three-photon resonance can be still observed by broadening its
linewidth, increasing the ratio $\Omega_B/\Omega_R$. Numerical
calculations show that more than 97 \% occupation of state
$|Q\rangle$ can be reached with lasers of bandwidth 10 kHz (HWHM)
under the three-photon resonance condition. For the same
parameters, the effect of the laser bandwidth at two+one-photon
resonance leads to a reduction of the population by only 0.2\% with
respect to the ideal case.

In conclusion, we have shown that the weak perturbation of a
$\Lambda$-system, achieved by coupling to a metastable state,
gives rise to novel dynamics. We have provided simple pictures for
understanding them. These can find applications for high-precision
measurement, for instance for metrology in the spirit
of~\cite{santra05,hong05}, quantum state preparation and
manipulation like for instance in transistors for single atoms as
in~\cite{micheli04}.

\begin{acknowledgments} C.C. thanks Jean Dalibard for very helpful
and stimulating discussions. G.M. acknowledges discussions with
Ramon Corbal\'an and the kind hospitality of the Laboratoire
PIIM-CIML at the Universit\'{e} de Provence under a visiting
professor grant. This work was partly supported by the French
Minist\`ere des affaires \'etrang\`eres (Picasso 09133XH), by the
Spanish Ministerio de Educaci\'on y Ciencia (Acci\'{o}n Integrada;
LACSMY project, FIS2004-05830; Ramon-y-Cajal Fellowship; QLIQS
project, FIS2005-08257) and by the European Commission (SCALA
Integrated Project, contract No. 015714). \end{acknowledgments}

\begin{appendix}

\section{Oscillatory motion of the trapped atom}

In the Lamb-Dicke regime, we decompose $$H_I(t)=H_I^{(0)}+{\rm
e}^{{\rm i}\nu t}H_I^{+}+ {\rm e}^{-{\rm i}\nu t}H_I^{-},$$ with
\begin{eqnarray*} H_I^{(0)}
&=&\frac{\hbar\Omega_B}{2}|P\rangle\langle
S|+\frac{\hbar\Omega_R}{2}|P\rangle\langle
D|+\frac{\hbar\Omega_C}{2}|Q\rangle\langle S|\\
& &+{\rm H.c} \\
H_I^{\pm} &=&{\rm i}\eta_B\frac{\hbar\Omega_B}{2}|P\rangle\langle
S|+{\rm i}\eta_R\frac{\hbar\Omega_R}{2}|P\rangle\langle D|\\ &
&+{\rm i}\eta_C\frac{\hbar\Omega_C}{2}|Q\rangle\langle S|+{\rm
H.c}. \end{eqnarray*} The solution can be found using a Floquet
ansatz for the density matrix, hence writing \begin{equation}
\rho=\sum_{n=-\infty}^{\infty}\rho^{(n)}{\rm e}^{{\rm i}n\nu t}
\end{equation} Substituting into the master equation, we find the
coupled equations
 \begin{eqnarray}\label{Master:Floquet}
\frac{\partial}{\partial t}\rho^{(n)}&=&-{\rm
i}n\nu\rho^{(n)}-\frac{\rm
i}{\hbar}[H_0+H_I^{(0)},\rho^{(n)}]\\
& &-\frac{\rm i}{\hbar}[H_I^{+},\rho^{(n-1)}]-\frac{\rm
i}{\hbar}[H_I^{-},\rho^{(n+1)}]+{\cal L}\rho^{(n)} \nonumber
 \end{eqnarray}
 which have been obtained by neglecting higher orders in the
 Lamb-Dicke expansion.

\end{appendix}


\end{document}